\documentclass[aps,prb,reprint,floatfix]{revtex4-1}

\usepackage{amsmath,amsfonts,amssymb,graphicx,braket}
\usepackage[]{units}
\usepackage{hyperref}
\hypersetup{
    colorlinks=true,
    linkcolor=blue,
    filecolor=blue,      
    urlcolor=blue,
    citecolor=blue
}

\begin{document}
\title{Electric Dipole Spin Resonance of 2D Semiconductor Spin Qubits}
\author{Matthew Brooks}
\email{matthew.brooks@uni-konstanz.de}
\author{Guido Burkard}
\affiliation{Department of Physics, University of Konstanz, D-78464 Konstanz, Germany}

\begin{abstract}
Monolayer transition metal dichalcogenides (TMDs) offer a novel two-dimensional platform for semiconductor devices. One such application, whereby the added low dimensional crystal physics (i.e. optical spin selection rules) may prove TMDs a competitive candidate, are quantum dots as qubits. The band structure of TMD monolayers offers a number of different degrees of freedom and combinations thereof as potential qubit bases, primarily electron spin, valley isospin and the combination of the two due to the strong spin orbit coupling known as a Kramers qubit. Pure spin qubits in monolayer Mo$X_2$ (where $X=$ S or Se) can be achieved by energetically isolating a single valley and tuning to a spin degenerate regime within that valley by a combination of a sufficiently small quantum dot radius and large perpendicular magnetic field. Within such a TMD spin qubit, we theoretically analyse single qubit rotations induced by electric dipole spin resonance. We employ a rotating wave approximation within a second order time dependent Schrieffer-Wolf effective Hamiltonian to derive analytic expressions for the Rabi frequency of single qubit oscilations, and optimise the mechanism or the parameters to show oscilations up to $\unit[250]{MHz}$.
\end{abstract}
\maketitle

\section{Introduction}
 \label{sec:Intro}
 
 Transition metal dichalcogenides (TMDs) are graphite-like indirect band-gap semiconductors in bulk, that when isolated down to the monolayer (ML) limit become two-dimensional visible range direct band-gap semiconductors, with a hexagonal crystal lattice structure\cite{wang2012electronics,kumar2012electronic,chhowalla2013chemistry,zhang2014direct,kormanyos2015k}. The combination of optically addressable electron spin and valley isospin degrees of freedom\cite{xiao2012coupled,xu2014spin} and strong spin-orbit coupling\cite{zhu2011giant,wang2015spin} within a mechanically flexible ML\cite{ccakir2014mechanical,palacios2017large} which may be stacked with other ML materials as part of the van der Waals (vdW) heterostructure engineering architecture\cite{geim2013van,withers2015light,zhong2017van}, has allowed for TMDs to be a viable and desirable host for quantum technologies. Quantum dots (QDs)\cite{pisoni2018gate}, single-photon emitters\cite{palacios2017large,branny2017deterministic,kern2016nanoscale}, gate defined nano-wires\cite{lin2014flexible,klinovaja2013spintronics}, topological materials\cite{fei2017edge,ma2016two}, ML superconductors\cite{xi2016ising,hsu2017topological} as well as spin-\cite{morpurgo2013spintronics,ghiasi2017large} and valley-tronics\cite{schaibley2016valleytronics,luo2017opto} have all been proposed or demonstrated with TMD MLs.
 
 \begin{figure}[!hb]
     \includegraphics[width=\linewidth]{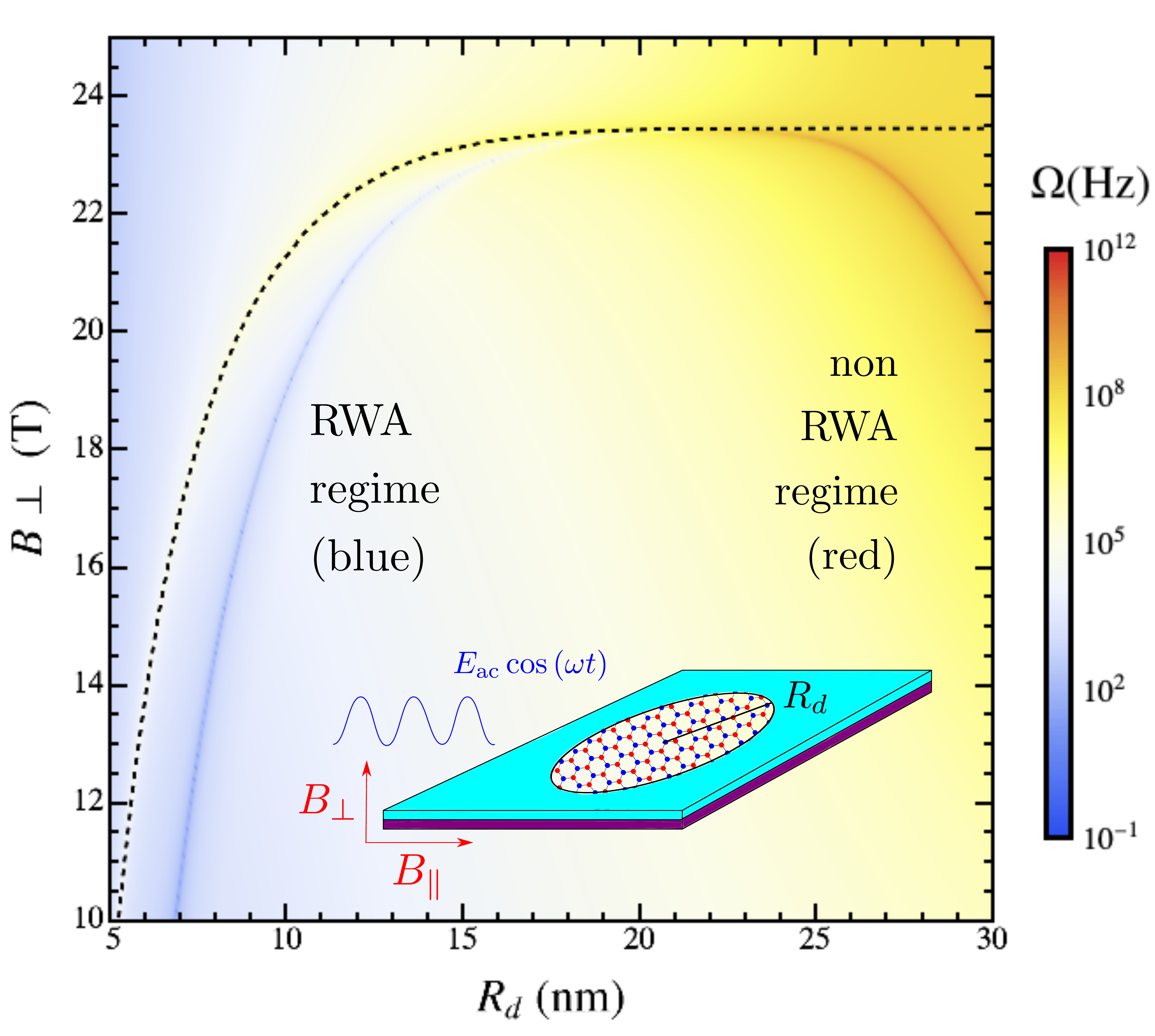}
     \caption{Rabi frequency $\Omega$ of a MoS$_2$ QD in dependance of the dot radius $R_{\text{d}}$ and out-of-plane magnetic field $B_\perp$ where $E_{\textrm{ac}}=\unit[10^{-2}]{V/nm}$ and $B_{\parallel}=\unit[1]{T}$. The black dashed line gives the points of spin degeneracy in the ground states of the $K'$ valley. Note that the region where the RWA is valid is where the freqencies calculated off-resonantly from the spin degeneracy line are small (blue), whilst the region where the maximum $\Omega$ deviates from the spin degeneracy line is where the RWA breaks down. Inset: diagram of the setup considered in this work of a gated TMD QD of radius $R_{\text{d}}$ (purple representing the TMD ML and cyan representing the top gate), exposed to a static out-of-plane magnetic field $B_\perp$, in-plane magnetic field $B_\parallel$ and an in plane AC-electric field $E_{\textrm{ac}}$.}
     \label{fig:diagram}
 \end{figure}
 
 Chemically, the semiconducting TMD MLs consist of $MX_2$ where $M=$ Mo or W and $X=$ S or Se, where the $M$ atomic layer is sandwiched between two $X$ atomic layers\cite{wang2012electronics,kumar2012electronic,chhowalla2013chemistry,zhang2014direct}, with broken inversion symmetry\cite{wang2012electronics,xu2014spin,yin2014edge}, and an $M-X$ alternating hexagonal structure in the plane of the ML\cite{wang2012electronics,chhowalla2013chemistry,zeng2015optical}. The $M$ atoms introduce strong spin-orbit coupling\cite{zhu2011giant,wang2015spin}, which with the broken inversion symmetry gives rise to spin-split conduction and valence bands\cite{kosmider2013large,kormanyos2013monolayer,xiao2012coupled}. Under an out-of-plane magnetic field, the splitting between the spin states in the conduction band is shifted due to both a spin- and valley-Zeeman effect\cite{srivastava2015valley,wang2017valley,chu2014valley,chu2014valley,lyons2018valley} introduced by a significant Berry curvature at the band-edges\cite{srivastava2015valley,yu2014dirac,kormanyos2018tunable}. Additionally, the Berry curvature allows for optically addressable spin-valley states by correctly applied circularly-polarised light\cite{yu2014dirac,xiao2012coupled}.

 QDs in TMD monolayers have been demonstrated by a number of differenct methods. Electrostatic gating\cite{pisoni2018gate}, strain\cite{palacios2017large,branny2017deterministic} and lattice defects\cite{brotons2018coulomb} have all been shown to achieve 0-dimensional behaviour in TMD monolayers. Strain and electrostatic gating however exhibit the most promise for QDs for quantum information purposes\cite{kormanyos2014spin}, and a number of different methods of implementing a qubit in a TMD QD have been proposed including spin-valley Kramers qubits\cite{szechenyi2018impurity}, in which one and two qubit gates have been proposed\cite{szechenyi2018impurity,david2018effective}, and pure-spin qubits\cite{brooks2017spin}. Pure-spin qubits, were shown to be achieveable by tuning a combination of the QD radius and the out of plane magnetic field such that, within one valley, a near-spin-degeneracy is reached. The magnetic field required to do so is high ($\sim\unit[20-30]{T}$) when considering only the natural spin- and valley-Zeeman contributions of the ML. However, as previously mentioned, one of the benefits of 2D semiconductors is the access to vdW heterostructure engineering. Thus,  it has been shown that by layering TMDs with magnetic monolayers such as CrI$_3$ and EuS, local time reversal symmetery violation in the TMD occurs, significantly enhancing the valley-Zeeman effect obsered in the TMD\cite{zhao2017enhanced,zhong2017van,seyler2018valley,huang2017layer}. A similar result may also be achieved with doping\cite{wang2018strongly}. The modularity of vdW heterostructure devices, along with an optically initialisable spin state, makes TMD QDs a strong contender to more conventional bulk semiconductor qubit realisations.

 Towards building a 2D quantum processor, the next step, after realising a qubit, is a scheme for single qubit gates, i.e. a reliable method of single-qubit state initilasation and control. In this work, we demonstrate that electric-dipole spin resonance (EDSR) may be achieved in TMD pure-spin qubits. EDSR requires the coupling of the qubit spin states to an external AC-electric field\cite{russ2017three,golovach2006electric}, which drives rotations between the spin states, such that ideally microwave pulses can be used to perform the desired single qubit gate. This has been theoretically shown to be achievable in TMD QDs adopting a Kramers qubit architechure, with the aid of an additional lattice defect to mix the valley states\cite{szechenyi2018impurity}. We show that in a valley-polarised pure-spin qubit architechure, EDSR is achievable and with some parameter optimisation (dot radius, magnetic fields etc.) oscilations of the qubit in the $\sim\unit[100]{MHz}$ regime are feasible.
 
 This paper is structured as follows, firstly, in Sec.~\ref{sec:Model}, the TMD QD Hamiltonian is given and the studied material type and parameter regime for the pure-spin qubit architechures is detailed. Then, in Sec.~\ref{sec:EDSR}, the EDSR mechanism is introduced in detail, giving all relevant matrix elements, as well as an effective qubit Hamiltanian given by a time dependent Schrieffer-Wolff transformation. Thirdly, in Sec.~\ref{sec:RWA}, the rotating wave approximation is applied to derive expressions for the Rabi frequency in the rotating frame. This is followed in Sec.~\ref{sec:optim} by an analysis of the relevant parameters of the system to maximise the qubit frequency. Lastly, in Sec.~\ref{sec:Discussion}, a discussion and comparision of this architechure with other known architectures is provided.
 
\section{Monolayer TMD Quantum Dots}
 \label{sec:Model}
  In this work we assume an electrostatic-gate defined QD in a TMD monolayer. With the appropriate selection of the TMD type, and a sufficiently large external magnetic field, it has been shown that the spin-valley locking may be overcome to provide a host for a valley polarised pure spin qubit\cite{brooks2017spin}.
\subsection{Effective Hamiltonian}
 \label{sec:effective}
The energy levels of a single electron in a TMD quantum dot in a perpendicular magnetic field ($B_\perp$) at the $K$ or $K'$ valleys may be obtained by solving the effective low energy Hamiltonian\cite{kormanyos2014spin,brooks2017spin} 
\begin{equation}
	\begin{split}
	   H_{B_\perp}^{\tau,s}= \hbar\omega_c^{\tau,s}\alpha_+\alpha_-+\tau s \frac{\Delta_{\text{cb}}}{2}+\frac{1+\tau}{2}\frac{B_\perp}{|B_\perp|}\hbar\omega_c^{\tau,s}   \\ 
	    +\frac{1}{2}(\tau g_{\text{vl}}+s g_{\text{sp}})\mu_B B_\perp& 
	    \end{split}
	    \label{eq:BPerp_Hamiltonian}
\end{equation}
where $\tau=\pm1$ is the valley index with $1(-1)\equiv K(K')$, $s=\pm1$ is the spin index with $1(-1)\equiv \uparrow(\downarrow)$, $\omega_c^{\tau,s}$ is the spin-valley dependent cyclotron frequency, $\Delta_{\text{cb}}$ is the spin-orbit splitting in the conduction band of the TMD, $g_{\text{vl}}$ and $g_{\text{sp}}$ are the valley and spin out of plane g-factors respectively and $\mu_B$ is the Bohr magneton. The spin-valley dependance of $\omega_c^{\tau,s}$ is due to the spin-valley dependance of the effective mass at the band edges given as $1/m_{\text{eff}}^{\tau,s}=1/m_{el}^0-\tau s/\delta m_{\text{eff}}$ where $\delta m_{\text{eff}}$ is contingent on the TMD type. The modified wavenumber operators $\alpha_{\pm}$ are $\alpha_{\pm}=\mp i l_Bq_{\pm}/\sqrt{2}$ where $l_B=\sqrt{\hbar/eB_\perp}$ is the magnetic length and $q_\pm=q_x\pm i q_y$ where $q_k=-i\partial_k$. The potential of the QD is assumed to be an infinite square well of radius $R_{\text{d}}$, which is reasonable when assuming the electrostatic gates of the dot to be contacted to or seperated by 1-2 layers 2D dielectric hexagonal boron nitride\cite{lee2015highly,pisoni2018gate}. Thus the quantum dot levels as a function of $B_\perp$ and $R_{\text{d}}$ are given as

\begin{equation}
    \tilde{\varepsilon}^{\tau,s}_{n,l}= \varepsilon^{\tau,s}_{n,l}+\tau s \frac{\Delta_{\text{cb}}}{2}
\label{eq:BPerp_total_Energy}
\end{equation} 

\noindent where

\begin{equation}
	\begin{split}
		\varepsilon^{\tau,s}_{n,l}= \hbar\omega_c^{\tau,s}\left(\frac{1+\tau}{2}\frac{B_\perp}{|B_\perp|}+          \frac{|l|+l}{2}-\gamma_{n,l}\right)& \\ +\frac{1}{2}(\tau g_{\text{vl}}+sg_{\text{sp}} )\mu_BB_\perp&. 
	\end{split}
\label{eq:BPerp_dot_Energy}
\end{equation} 

\noindent Here $\gamma_{n,l}$ is the $n^{\text{th}}$ solution to $M(\gamma_{n,l},|l|+1,R_{\text{d}}^2/2l_B^2)=0$, where $M(a,b,c)$ is the confluent hypergeometric function of the first kind, given by the hard-wall boundary conditon to Eq.~(\ref{eq:BPerp_Hamiltonian}). 

\subsection{Single dot spin qubit}
 \label{sec:spinQb}
The spin-valley locking due to spin-orbit coupling and crystal symetries can be shown to be overcome, resulting in a pure spin qubit\cite{brooks2017spin} with a TMD QD as opposed to a spin-valley Kramers' qubit\cite{szechenyi2018impurity}. By selecting the appropriate TMD type, dot size and perpendicular magnetic field a regime where $\varepsilon^{K(K'),\uparrow}_{n,l}=\varepsilon^{K(K'),\downarrow}_{n,l}$ may be achieved. MoS$_2$ is the semiconducting TMD monolayer with the smallest zero field spin splitting in the conduction band $\Delta_{\text{cb}}$ and a $\delta m_{\text{eff}}$ such that the condition $\varepsilon^{K',\uparrow}_{1,0}=\varepsilon^{K',\downarrow}_{1,0}$ may be achieved for $B_\perp\approx\unit[16]{T}$ in the first excited state ($n=1$, $l=-1$) and $B_\perp\approx\unit[21]{T}$ in the ground state ($n=1$, $l=0$) assuming $R_d\approx\unit[10]{nm}$. Assuming that the QD is charged by a valley polarised source, either optically or by valley polarised leads, a pure spin qubit in an MoS$_2$ monolayer gated quantum dot may be realised.

\section{Electric Dipole Spin Resonance}
 \label{sec:EDSR}
\subsection{External Influences}
  \label{sec:Fields}
    
  To achieve control over the qubit spin states, two additional ingredients to the spin-orbit interaction inherent in the crystal are needed; a spin-mixing interaction and a driving field. These are achieved by subjecting the QD to a static in-plane magnetic field and AC in-plane electic field.
  
  The Hamiltonian describing an in-plane magnetic field along the $x$-direction is given as
  \begin{equation}
  	H_{B_{\parallel}}=\frac{1}{2} \mu_B g_{\parallel} B_{\parallel}s_x
	\label{eq:Bx}
  \end{equation} 
  
  \noindent where $g_{\parallel}$ is the in-plane g-factor, $B_{\parallel}$ is the in-plane magnetic field and $s_i$ where $i=(x,y,z)$ is the $i$\textsuperscript{th} spin Pauli matrix, i.e. $s_i=(\hbar/2)\sigma_i$. The in-plane g-factor is assumed in this work to be $g_{\parallel}=2$, as we assume a clean crystal sample. The out-of-plane g-factor $g_s$ is material dependent and given by the same 7-band $k\cdot p$ analysis used to derive the effective Hamiltonian Eq.~(\ref{eq:BPerp_Hamiltonian})\cite{kormanyos2014spin}.
  
  The real-space Hamiltonian of an AC-electric driving field along the $x$-direction is given as 
  \begin{equation}
  	\tilde{H}_{ac}=exE_{ac}\cos{\omega t}
	\label{eq:HacTilde}
  \end{equation} 
  
  \noindent where $e$ is the elementary charge, $E_{ac}$ and $\omega$ denote the field strength and frequency of the AC-field and $t$ is time. In the orbital basis this can be rewritten as approximately (see App.~\ref{sec:dipole}):
  
  \begin{equation} 	
	  H_{ac}=\sigma_x \frac{eE_{\text{ac}}R_{\text{d}}\cos(\omega t)}{2\sqrt{2}}
	\label{eq:Hac}
  \end{equation}
  
  \noindent where $\sigma_i$ is the $i$\textsuperscript{th} orbital Pauli matrix. From these matrix elements, the full Hamiltonian for ESDR in TMD QDs may be written.
  
\subsection{4 $\times$ 4 valley-polarised Hamiltonian}
  \label{sec:4x4}

Due to our choice of material and $B_\perp$ direction (positive along the $z$-axis), the valley in which the spin qubit is achieved is the $K'$. From all the elements collected in Sec.~\ref{sec:Model} and~\ref{sec:Fields}, the full Hamiltonian of the valley-polarised TMD dot with an in-plane magnetic field and AC-electic field is
\begin{widetext}
    \begin{equation}
  	  	H_{K'}=
  		\frac{1}{2}\left(\begin{array}{cccc}
  			2\varepsilon^{K',\uparrow}_{1,0}-\Delta_{cb} & \mu_B g_\parallel B_\parallel  & \frac{e E_{\text{ac}}R_{\text{d}}\cos(\omega t)}{\sqrt{2}} & 0 \\
  			\mu_B g_\parallel B_\parallel& 2\varepsilon^{K',\downarrow}_{1,0}+\Delta_{cb} & 0 & \frac{e E_{\text{ac}}R_{\text{d}}\cos(\omega t)}{\sqrt{2}} \\
  			\frac{e E_{\text{ac}}R_{\text{d}}\cos(\omega t)}{\sqrt{2}} & 0 & 2\varepsilon^{K',\uparrow}_{1,-1} -\Delta_{cb}  & \mu_B g_\parallel B_\parallel\\
  			0 & \frac{e E_{\text{ac}}R_{\text{d}}\cos(\omega t)}{\sqrt{2}} & \mu_B g_\parallel B_\parallel & 2\varepsilon^{K',\downarrow}_{1,-1} +\Delta_{cb}
  		\end{array}\right)			
  	  \label{eq:4x4}
    \end{equation}
\end{widetext}

\noindent for the qubit basis and the first excited orbital spin states ($\{|l=0,K',\uparrow\rangle,|l=0,K',\downarrow\rangle,|l=-1,K',\uparrow\rangle,|l=-1,K',\downarrow\rangle\}$) to which the qubit couples by the driving field. From this, an approximate $2\times2$ time dependent qubit Hamiltanian may be derived. 
  
\subsection{Time dependent Schrieffer-Wolff transformation}
 \label{sec:TDSWT}
 
	A second order time dependent Schrieffer-Wolff transformation (TDSWT) is employed to isolate a time dependent effective qubit Hamiltonian\cite{romhanyi2015subharmonic} (for a complete derevation see App.~\ref{sec:AppTDSW}). The relevant terms of the transformation are

	\begin{equation}
		H_{\textrm{EDSR}}(t)=\tilde{\mathcal{H}}^{(0)}+\tilde{\mathcal{H}}^{(1)}+\tilde{\mathcal{H}}^{(2)}(t)
		\label{eq:TDSW_Simple}
	\end{equation}

	\noindent where 
	
	\begin{subequations}
		\begin{equation}
			\tilde{\mathcal{H}}^{(0)}=\sum_{s,l}\tilde{\varepsilon_{1,l}^{\tau,s}}\ket{s,l}\bra{s,l}
			\label{eq:TDSW_Term_A}
		\end{equation}
		\begin{equation}
			\tilde{\mathcal{H}}^{(1)}=\frac{\mu_B g_\parallel B_\parallel}{2}s_x
			\label{eq:TDSW_Term_B}
		\end{equation}
		\begin{equation}
			\tilde{\mathcal{H}}(t)^{(2)}=\frac{E_{\text{ac}}^2R_{\text{d}}^2[1+\cos(2\omega t)]}{36 \hbar \omega_{s,s}^{0,-1}}\sigma_z
			\label{eq:TDSW_Term_C}
		\end{equation}
	\end{subequations}

	\noindent where $\omega_{s,s'}^{l,l'}$ is the energy difference between the two QD levels $\varepsilon^{K',s}_{1,l}$ and $\varepsilon^{K',s'}_{1,l'}$ expressed as an angualar frequency such that, for example $\varepsilon^{K',\uparrow}_{1,0}-\varepsilon^{K',\downarrow}_{1,-1}=\hbar \omega_{\uparrow,\downarrow}^{0,-1}$. The small parameters for the TDSWT are the electric field strength $e E_{\text{ac}}R_{\text{d}}/\hbar\omega_{\uparrow,\downarrow}^{0,0}\ll1$ and in plane magnetic field strength $\mu_B g_\parallel B_\parallel/2\hbar\omega_{\uparrow,\downarrow}^{0,0}\ll1$. Accordingly, Eq.~(\ref{eq:TDSW_Simple}) leads to a block diagonal Hamiltonian for which the relevant time dependent qubit basis portion may be extracted as

	\begin{equation}
		\begin{split}
			&H_{\textrm{EDSR}}(t)=\\&\begin{pmatrix} 
				\varepsilon^{K',\uparrow}_{1,0}+\frac{e^2E_{\text{ac}}^2R_{\text{d}}^2[1+\cos(2\omega t)]}{16 \hbar \omega_{\uparrow,\uparrow}^{0,-1}} & \frac{\mu_B g_\parallel B_\parallel}{2} \\
				\frac{\mu_B g_\parallel B_\parallel}{2} & \varepsilon^{K',\downarrow}_{1,0}+\frac{e^2E_{\text{ac}}^2R_{\text{d}}^2[1+\cos(2\omega t)]}{16 \hbar \omega_{\downarrow,\downarrow}^{0,-1}}
			\end{pmatrix}.
		\end{split}
		\label{eq:qubitHamiltonian}
	\end{equation}

\section{Rabi oscilations}
 \label{sec:RWA}
 
 From the time-dependent qubit Hamiltonian given in Eq.~(\ref{eq:qubitHamiltonian}), a transformation into the rotating basis may be performed and the rotating-wave approximation applied to derive the Rabi-oscillation frequency in the rotating frame as
 \begin{widetext}
    \begin{equation}
    	\tilde{\Omega}=\frac{3\mu_B g_\parallel B_\parallel e^2 E_{\text{ac}}^2 R_{\text{d}}^2 \left(\omega_{\downarrow,\downarrow}^{0,-1}-\omega_{\uparrow,\uparrow}^{0,-1}\right)}{4\sqrt{ \left(36 \hbar \mu_{B} g_{\parallel} B_{\parallel}\omega_{\uparrow,\uparrow}^{0,-1}\omega_{\downarrow,\downarrow}^{0,-1}\right)^2 +\left(e^2E_{\text{ac}}^2R_{\text{d}}^2\left[\omega_{\downarrow,\downarrow}^{0,-1}-\omega_{\uparrow,\uparrow}^{0,-1}\right]-36 \hbar^2\omega_{\uparrow,\downarrow}^{0,0}\omega_{\uparrow,\uparrow}^{0,-1}\omega_{\downarrow,\downarrow}^{0,-1}\right)^2}}.
		\label{eq:fullRABI}
    \end{equation}
 \end{widetext}
 
 \noindent Note that in this form, the implicit dependance of the Rabi frequency $\tilde{\Omega}$ on $B_\perp$ is within all the $\omega_{s,s'}^{l,l'}(B_\perp)$ frequencies while the depenance of $\tilde{\Omega}$ on the spin-orbit splitting of the conduction band $\Delta_{\textrm{cb}}$ is within $\omega_{\uparrow,\uparrow}^{0,-1}(B_\perp,\Delta_{\textrm{cb}})$ and $\omega_{\downarrow,\downarrow}^{0,-1}(B_\perp,\Delta_{\textrm{cb}})$. The difference between the two however, present in the numerator of Eq.~(\ref{eq:fullRABI}) is not dependent on the spin-orbit splitting. Note that, as the spin splitting due to the spin orbit interaction is decreased, so too is the maximum Rabi frequency achievable, and as $\Delta_{\textrm{cb}}\rightarrow0$ the in-plane magnetic field small parameter condition of the TDSWT is violated and all of the calculations made up to this point are no longer valid.  
 
 A further simplification of Eq.~(\ref{eq:fullRABI}) may be given as its dominant term 
 
 \begin{equation}
	 	\Omega=\frac{\mu_B g_\parallel B_\parallel e^2E_{\text{ac}}^2R_{\text{d}}^2 \left(\varepsilon_{1,-1}^{K',\uparrow}-\varepsilon_{1,-1}^{K',\downarrow}\right)}{48\Delta_{\text{cb}}\left[\varepsilon_{1,0}^{K',\uparrow}-\varepsilon_{1,-1}^{K',\uparrow}\right]\left[\varepsilon_{1,0}^{K',\downarrow}-\varepsilon_{1,-1}^{K',\downarrow}\right]\hbar^2} 
	\label{eq:reducedRABI}
 \end{equation}
 
 \noindent assuming $\varepsilon_{1,0}^{K',\uparrow}\approx\varepsilon_{1,0}^{K',\downarrow}$, i.e. operating at the spin qubit regime. The physics of the terms dropped from (\ref{eq:fullRABI}) to give (\ref{eq:reducedRABI}) are apparent from the following expansion
 
 \begin{equation}
 	\tilde{\Omega}=\Omega(1+\delta_1+\delta_2+\dots)
	\label{eq:expansion}
 \end{equation}
 
 \noindent where
\begin{subequations}
	\begin{equation}
		\begin{split}
		\delta_1=\frac{e^2E_{\text{ac}}^2R_{\text{d}}^2\left[\omega_{\downarrow,\downarrow}^{0,-1}-\omega_{\uparrow,\uparrow}^{0,-1}\right]}{36 \hbar^2\omega_{\uparrow,\downarrow}^{0,0}\omega_{\uparrow,\uparrow}^{0,-1}\omega_{\downarrow,\downarrow}^{0,-1}}&\\\times\left(1+\frac{e^2E_{\text{ac}}^2R_{\text{d}}^2\left[\omega_{\downarrow,\downarrow}^{0,-1}-\omega_{\uparrow,\uparrow}^{0,-1}\right]}{72 \hbar^2 \omega_{\uparrow,\downarrow}^{0,0}\omega_{\uparrow,\uparrow}^{0,-1}\omega_{\downarrow,\downarrow}^{0,-1}}\right)&
		\end{split}
		\label{eq:higher_order_A}
	\end{equation}
	
	\begin{equation}
		\delta_2=\frac{\left(\mu_{B} g_{\parallel} B_{\parallel}\right)^2}{2\left(\hbar\omega_{\uparrow,\downarrow}^{0,0}\right)^2}.
		\label{eq:higher_order_B}
	\end{equation}
\end{subequations}
	
\noindent From this, $\delta_1$ can be reasoned as a shift due to the AC stark effect as it is a perturbation in a higher order of $E_{\textrm{ac}}$ and $\delta_2$ is the plane Zeeman shift due to $B_{\parallel}$. From this form of the Rabi frequency, the effect of the EDSR fields may be probed.

 \begin{figure}[!ht]
     \includegraphics[width=\linewidth]{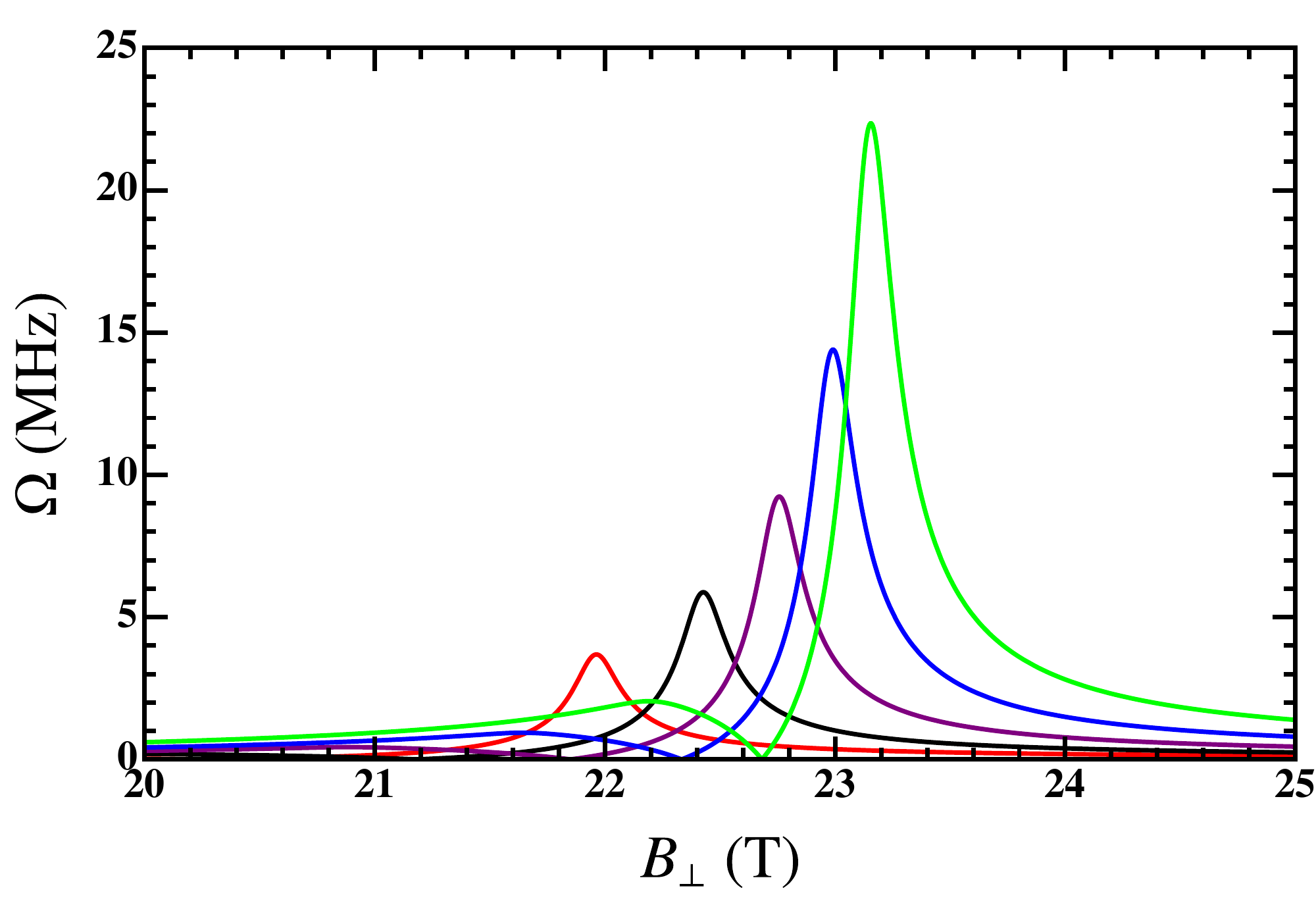}
     \caption{The out-of-plane magnetic field $B_\perp$ dependance of the Rabi frequency $\Omega$ for MoS$_2$ QDs with $R_d=\unit[11]{nm}$ (red), $\unit[12]{nm}$ (black), $\unit[13]{nm}$ (purple), $\unit[14]{nm}$ (blue) and $\unit[15]{nm}$ (green), with $E_{\textrm{ac}}=\unit[10^{-2}]{V/nm}$ and $B_\parallel=\unit[50]{mT}$.}
     \label{fig:BPerp}
 \end{figure}
 
 \begin{figure}[!ht]
     \includegraphics[width=\linewidth]{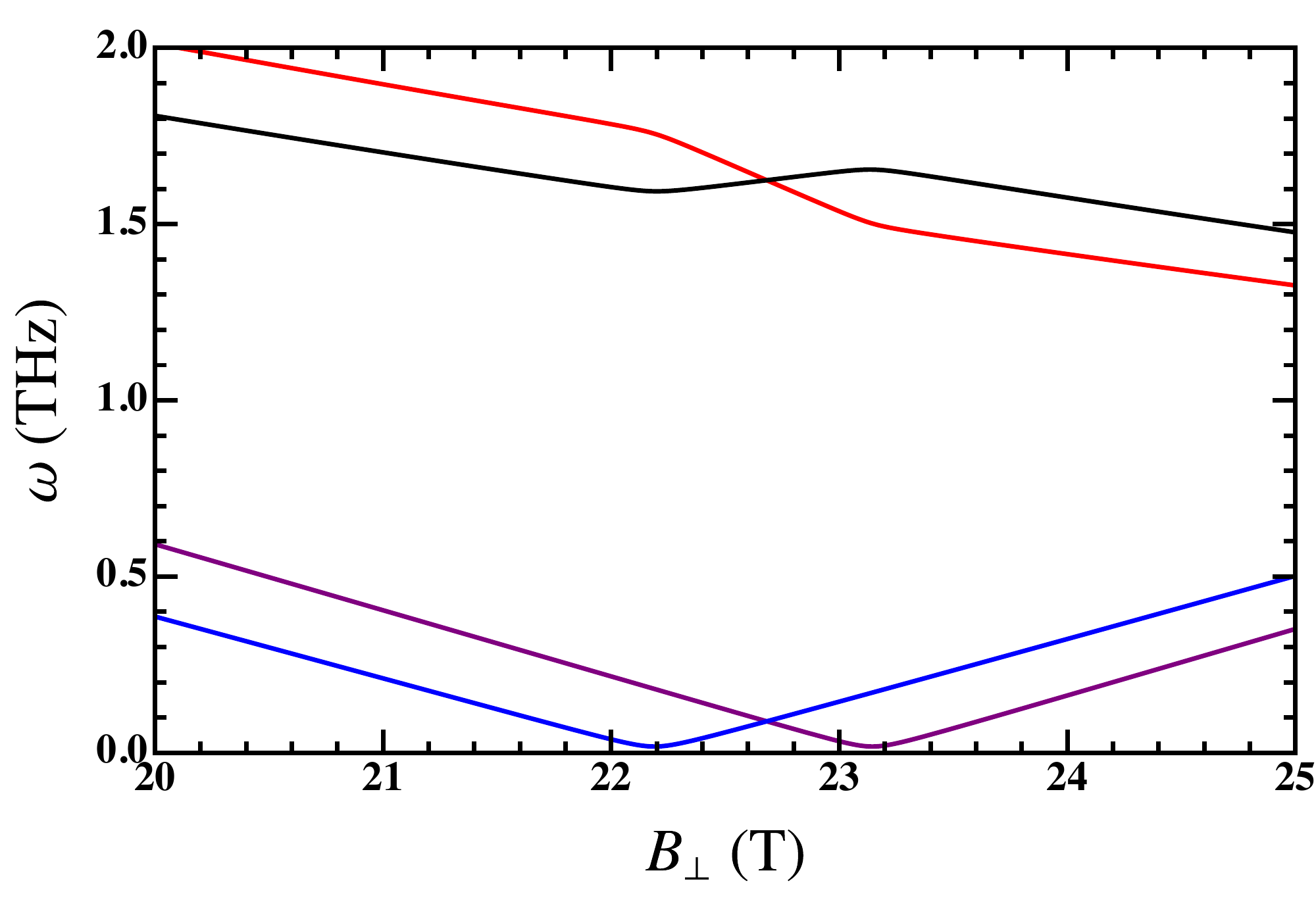}
     \caption{The out-of-plane magnetic field $B_\perp$ dependance of the QD level splittings expressed as angular frequencies $\omega_{\downarrow,\downarrow}^{0,-1}$ (red), $\omega_{\uparrow,\uparrow}^{0,-1}$ (black), $\omega_{\uparrow,\downarrow}^{0,0}$ (purple) and $\omega_{\uparrow,\downarrow}^{-1,-1}$ (blue), for MoS$_2$ QDs of with $E_{\textrm{ac}}=\unit[10^{-2}]{V/nm}$, $B_\parallel=\unit[50]{mT}$ and $R_{\text{d}}=\unit[15]{nm}$.}
     \label{fig:omega_split}
 \end{figure}

\begin{figure}[!ht]
     \includegraphics[width=\linewidth]{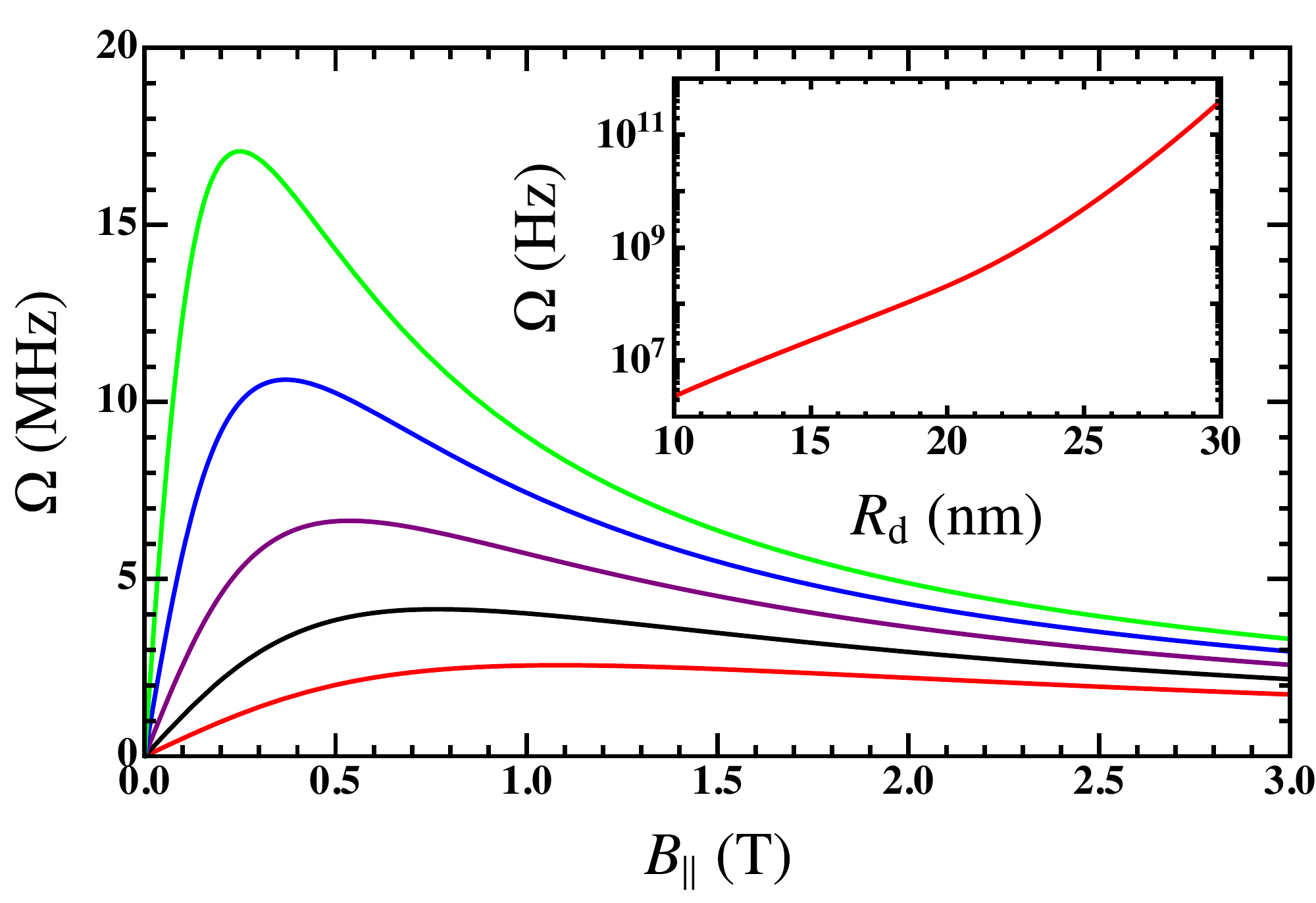}
     \caption{Rabi frequency on resonance for MoS$_2$ QDs with $R_d=\unit[11]{nm}$ (red), $\unit[12]{nm}$ (black), $\unit[13]{nm}$ (purple), $\unit[14]{nm}$ (blue) and $\unit[15]{nm}$ (green), and $E_{\textrm{ac}}=\unit[10^{-2}]{V/nm}$. Inset: Extracted maximum Rabi frequency $\Omega$ with dot radius $R_{\text{d}}$ for MoS$_2$ QDs with $E_{\textrm{ac}}=\unit[10^{-2}]{V/nm}$ and $B_{\parallel}=\unit[1]{T}$.}
     \label{fig:BParallel}
 \end{figure}

Firstly, the effect of the strength of the AC-electric field $E_{\textrm{ac}}$ is clearly quadratic. As such, this value shall be fixed at $\unit[10^{-2}]{V/nm}$, a reasonably achievable electric field amplitude that is consistent with the validity of the small parameter assumption in the following calculations. The effect of $B_\perp$ can be seen in both Fig.~\ref{fig:BPerp} and~\ref{fig:omega_split}. Fig.~\ref{fig:BPerp} shows the dependance of $\Omega$ on $B_\perp$ for a number of dot radii. There is a clear peak for each radius and clear minimum, where $\Omega\rightarrow 0$, at which $\omega_{\downarrow,\downarrow}^{0,-1}=\omega_{\uparrow,\uparrow}^{0,-1}$. The reason for this interference is clear in Fig.~\ref{fig:omega_split}. The avoided crossings for the qubit states and the orbitally excited states do not align with $B_\perp$, as such, there are values of $B_\perp$ that are after one avoided crossing and before the second. This manifests itself in Fig.~\ref{fig:omega_split} where each of the kinks in the gradient of the $\omega_{\downarrow,\downarrow}^{0,-1}$ and $\omega_{\uparrow,\uparrow}^{0,-1}$ lines occur at the avoided crossings. It is in between these two kinks that the destrutcive intereference is such that $\omega_{\downarrow,\downarrow}^{0,-1}=\omega_{\uparrow,\uparrow}^{0,-1}$ and $\Omega\rightarrow0$. The effect of $B_\parallel$ is also not fully apparent from Eq.~(\ref{eq:reducedRABI}). Of course, from the denominator as $B_\parallel\rightarrow 0$ so does $\Omega\rightarrow 0$, as there is no spin mixing mechanism at this limit, but the relationship between the two is not linear, as a wider avoided crossing can be detrimental to the rotation speed. As is seen in Fig.~\ref{fig:BParallel}, there is a clear peak in the achieveable $\Omega$ at $\sim\unit[50]{mT}$, followed by a plateau at $\sim\unit[1]{T}$, for a range of radii. 
 
\section{Optimal Operations}
 \label{sec:optim}
 
 Understanding in detail the effects of each of the contributing EDSR mechanisms on the derived single qubit rotational frequency now allows for an optimisation of the EDSR procedure. However, there is still one parameter with which the mechanism may be optimised, the dot radius. Fig.~\ref{fig:diagram} gives $\tilde{\Omega}$ in dependence of $R_{\text{d}}$ and $B_\perp$ at constant $E_{\textrm{ac}}$ and $B_\parallel$, showing a clear peak running along the spin-degeneracy line as well as the interference line under the peak. Note that here the full expression $\tilde{\Omega}$ is plotted as to demonstrate where the RWA starts to break down, as for $R_d\gtrsim\unit[22.5]{nm}$, the higher order terms deviate the peak from around the spin degeneracy point and the Rabi frequency divererges past the reaonsable range of the assumed driving frequency (microwave). The reduced form of the Rabi frequency $\Omega$ gives exactly the same result below this point, without showing the deviation at larger dot radii. The inset of Fig.~\ref{fig:BParallel} shows more explicitly the $R_{\text{d}}$ dependance of the maximum Rabi frequency achieveable when at a fixed $B_\parallel=\unit[1]{T}$. Here a near logarithmic increase in achieveable Rabi frequency. This trend is easily exploitable but comes with a significant cost in $B_\perp$.
 
\begin{figure}[!ht]
     \includegraphics[width=\linewidth]{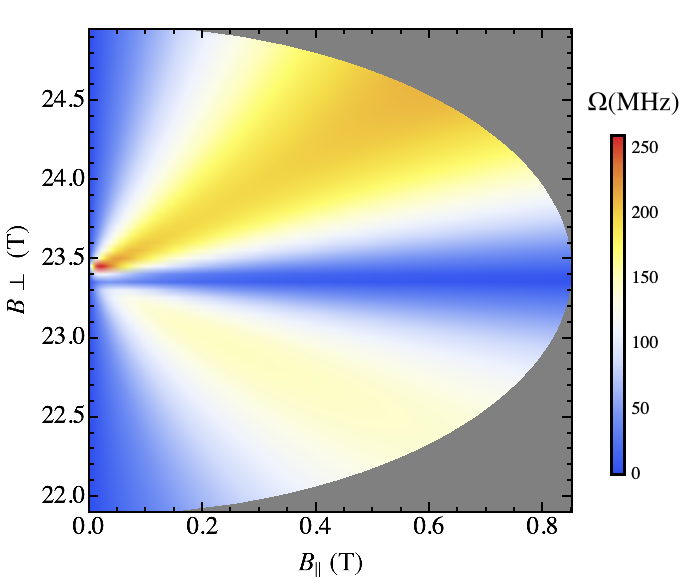}
     \caption{The out-of-plane $B_\perp$ and in-plane $B_\parallel$ magnetic field dependancies of the Rabi frequency $\Omega$ for an MoS$_2$ QD of radius $R_{\text{d}}=\unit[20]{nm}$ where $E_{\textrm{ac}}=\unit[10^{-2}]{V/nm}$ only within the microwave qubit detuning range.}
     \label{fig:optimised}
 \end{figure}

 As a proposal for an optimal operational regime, consider a dot of $R_{\text{d}}=\unit[20]{nm}$. To satisfy both the conditions of the RWA and experimental preferences, only the regime where the qubit detuning is within the microwave range $<\unit[300]{GHz}$ shall be considered. This is shown in Fig.~\ref{fig:optimised}, where a clear peak region at $B_\perp=\unit[23.5]{T}$ and $B_\parallel=\unit[20]{mT}$ can be seen. At this optimised point a very desirable Rabi frequency of $\sim\unit[250]{MHz}$ is reached. However, there is a band where Rabi frequencies $\sim\unit[100]{MHz}$ are attainable, allowing for less precise control of the magnetic fields to access a desirable frequency range.

\section{Discussion}
\label{sec:Discussion}
To implement a pure-spin qubit with fast single gate operations we find that a good choice consists of an MoS$_2$ QD of radius $R_{\text{d}}=\unit[20]{nm}$, in a external out-of-plane magnetic field $B_\perp=\unit[23.5]{T}$, in-plane magnetic field $B_\parallel=\unit[20]{mT}$ and a microwave frequency AC-electric field of strength $E_{\text{ac}}=\unit[10^{-2}]{V/nm}$. This allows for a Rabi frequency of $\Omega=\unit[250]{MHz}$. All of the assumed field parameters are within reasonable viability. The $B_\perp$ requirement is high, however this can be reasonably mitigated by vdW heterostructre engineering with magnetic monolayers. All calculations given assume the qubit is implemented in a free standing TMD ML, to give an upper limit on what would be experimentally required. Recent advances in vdW heterostructure engineering have shown that significant valley-Zeeman enhancement can be achieved by layering the TMD on a ML or low dimensional magnetic material\cite{zhao2017enhanced,zhong2017van,seyler2018valley}. Ideally, a vdW stack of hBN - CrI$_3$ or EuS - MoS$_2$ - hBN would be used to implement a TMD spin quantum processor. The purpose of the hBN is to protect the other MLs from degredation as well as improve the optical response of the TMD for state initialisation\cite{zhou2017probing,cadiz2017excitonic,scuri2018large}. 

The gate speed shown here is an order of magnitude faster within reasonable experimental limitations than has been shown in the alternative single dot approach to TMD qubits, the Kramers qubit\cite{szechenyi2018impurity}. This assumes a clean cystal, unlike the Kramers qubit that requires a defect to mix the valleys. While defects are currently inherent to TMD samples, they are usually undesirable, and in the proposed pure-spin qubit scheme offer a dephasing mechanism. However, the $K$-valley levels are higher in energy and become more energetically separated at lower $R_{\text{d}}$, therefore, some tradeoff between gate speed and stability can be made in the case of valley-mixing crystal defects. Additionally, there has recent significant progress in synthesising low defect rate monolayers by chemical as opposed to mechanical means\cite{pistunova2019transport}.

The $\sim\unit[100]{MHz}$ single gate rotations makes this 2D qubit implementation competitive with more conventional bulk semiconductor achitechures. Both GaAs and Si 2D electron gas gated single spin qubits have experimentally shown Rabi oscillations in the order of $\sim\unit[10]{MHz}$\cite{russ2017three,nadj2010spin,kawakami2014electrical}. However, in TMDs, these fast gate speeds are required as spin lifetimes have only been measured up to a few nanoseconds\cite{yang2015long}. This is however, expected to improve with the advent of cleaner crystal samples.The promise of similar to improved speeds attainable with the TMD device proposed here, in a flexible and optically active medium, further position 2D semiconductors as exciting novel materials for quantum device applications.




\section{Acknowledgements}
\label{ref:Acknowledgements}

We acknowledge helpful discussions with A. David, F. Ginzel, M. Russ and V. Shkolnikov and funding through both the European Union by way of the Marie Curie ITN Spin-Nano and the DFG through SFB 767.

\appendix
\section{Dipole Matrix}
\label{sec:dipole}

The dipole matrix elements represent the off-diagonal elements that in the case of this work couple the qubit states with the first excited orbital states. These are calculated as follows

\begin{equation}
	d_{nl,n'l'}=\langle\psi_{nl}|\tilde{H}_{ac}|\psi_{n'l'}\rangle
	\label{eq:dipole}
\end{equation}

\noindent where $\tilde{H}_{ac}$ is given by (\ref{eq:HacTilde}). Here the wavefunctions are derived from Eq.~(\ref{eq:BPerp_Hamiltonian}) as\cite{brooks2017spin}

\begin{equation}
	\psi_{n,l}=\mathcal{A}(\gamma_{n,l},\rho)e^{il\theta}\rho^{|l|/2} e^{-\rho/2}M (\gamma_{n,l},|l|+1,\rho)
\end{equation}

\noindent where $\mathcal{A}(\gamma_{n,l},\rho)$ is the normalising factor. Importantly for this work, the matrix element $\langle\psi_{n,l}|\tilde{H}_{ac}|\psi_{n,l}\rangle=0$ while $\langle\psi_{n,l}|\tilde{H}_{ac}|\psi_{n,l'}\rangle\neq0$ for $l\neq l'$. The value of these matrix elements can be calculated numerically. The corresponding matrix element is dependent on $B_\perp$ and $R_d$, however, we find that the dependance on $B_\perp$ is so slight ($<0.01\%$) that for this work we shall simply assume

\begin{equation}
	 \langle\psi_{0,1}|\tilde{H}_{ac}|\psi_{0,0}\rangle\approx\frac{eE_{ac}R_d}{2\sqrt{2}}.
\end{equation}

\section{Full TDSWT Derivation}
\label{sec:AppTDSW}
The time-dependent Schrieffer-Wolff transformation is a perturbative method to derive an effective block diagonal Hamiltonian $\tilde{\mathcal{H}}(t)$ from a dense Hamiltonian $\mathcal{H}(t)$ such as Eq.~(\ref{eq:4x4})\cite{romhanyi2015subharmonic}. We proceed by applying the unitary transformation $U(t)=e^{-S(t)}$, such that
\begin{equation}
	\tilde{\psi}(t)=e^{-S(t)}\psi(t),
\end{equation}
and, using the time-dependent Schr\"odinger equation, $-i\hbar\frac{\partial }{\partial t}\psi(t)+\mathcal{H}(t)\psi(t)=0$, leading to the transformed Hamiltonian
\begin{equation}
	\tilde{\mathcal{H}}(t)=e^{-S(t)}\mathcal{H}(t)e^{S(t)}+i\hbar\frac{\partial e^{-S(t)}}{\partial t}e^{S(t)}.
	\label{eq:TDSW}
\end{equation}

\noindent Here $S(t)$ is some block off-diagonal matrix. From this set up, a power-series expansion can then be applied which can be simplified to give
\begin{equation}
	\tilde{\mathcal{H}}(t)=\sum_{j=0}^{\infty}\frac{1}{j!}\left[\mathcal{H}(t),S(t)\right]^{(j)}-i\hbar\sum_{j=0}^{\infty}\frac{1}{(j+1)!}\left[\dot{S}(t),S(t)\right]^{(j)}
\end{equation}

\noindent where $\left[A,B\right]^{(0)}=A$ and $\left[A,B\right]^{(n+1)}=\left[\left[A,B\right]^{(n)},B\right]$. Here, $S(t)$ is solved for by assuming $\tilde{\mathcal{H}}(t)_{\text{off-diagonal}}=0$. At this point no approximation has been made. The approximation made to solve Eq.~(\ref{eq:TDSW}) such that $\tilde{\mathcal{H}}(t)_{\text{off-diagonal}}=0$ is a power-series expansion of the small parameters (in plane electric and magnetic fields) of the $S(t)$ matrix
\begin{equation}
	S(t)=S(t)^{(1)}+S(t)^{(2)}+S(t)^{(3)}+\dots 
\end{equation}
\noindent where $S(t)_n$ is the $n^{\text{th}}$ order of the power-series.

At this point, all the necessary definitions have been made to perform a general TDSWT, as such, now only a second order perturbation of the Eq.~(\ref{eq:4x4}) will be considered with small parameters are the electric field strength $e E_{\text{ac}}R_{\text{d}}/\hbar\omega_{\uparrow,\downarrow}^{0,0}(B_\perp)\ll1$ and in-plane magnetic field strength $\mu_B g_\parallel B_\parallel/\hbar\omega_{\uparrow,\downarrow}^{0,0}(B_\perp)\ll1$. The effective Hamiltonian with corrections up to second order is given by
\begin{equation}
	\tilde{\mathcal{H}}(t)=\tilde{\mathcal{H}}^{(0)}+\tilde{\mathcal{H}}^{(1)}+\tilde{\mathcal{H}}(t)^{(2)}.
	\label{eq:B6}
\end{equation}
\noindent From this, the expansions in $\tilde{\mathcal{H}}(t)$ can be solved from Eq.~(\ref{eq:TDSW}) as
\begin{subequations}
	\begin{equation}
		\tilde{\mathcal{H}}^{(0)}=\mathcal{H}_{0}
	\end{equation}
	\begin{equation}
		\tilde{\mathcal{H}}^{(1)}=\mathcal{H}_{1}
	\end{equation}
	\begin{equation}
		\tilde{\mathcal{H}}^{(2)}(t)=\frac{1}{2}\left[\mathcal{H}_{2}(t),S(t)^{(1)}\right].
	\end{equation}
\end{subequations}
\noindent Here $\mathcal{H}_{0}$ is the diagonal part of Eq.~(\ref{eq:TDSW}), $\mathcal{H}_{1}$ is the block diagonal part omitting the diagonal part of Eq.~(\ref{eq:TDSW}) and $\mathcal{H}_{2}(t)$ is the block off-diagonal part of Eq.~(\ref{eq:TDSW}), which for the case of the EDSR mechanism described translates as the QD levels $\mathcal{H}_{0}=\sum_{s,l}\varepsilon^{K',s}_{1,l}\ket{s,l}\bra{s,l}$, in-plane magetic field Eq.~(\ref{eq:Bx}) for $\mathcal{H}_{1}$ and AC-electric field maxtrix elements Eq.~(\ref{eq:Hac}) for $\mathcal{H}_{2}$. Only $S_1(t)$ needs to be solved for, which is done by applying the $\tilde{\mathcal{H}}(t)_{\text{off-diagonal}}=0$ condition giving
\begin{equation}
	\left[\mathcal{H}_{0},S(t)^{(1)}\right]=-\mathcal{H}_{2}.
\end{equation}

\noindent So finally, a block diagonal of the qubit and the excited orbital space may be approximated where the qubit space of Eq.~(\ref{eq:B6}) is given as Eq.~(\ref{eq:qubitHamiltonian}).



\bibliography{EDSRBibliography}

\end{document}